\documentclass[prl,preprint,showpacs]{revtex4}
\usepackage{graphicx}
\usepackage{dcolumn}
\usepackage{amsmath}

\begin{document}

\newcommand \bfig {\begin{figure}}
\newcommand \efig {\end{figure}}

\title{Cooperative Jahn-Teller
transition and resonant x-ray scattering in thin film ${\rm
LaMnO_3}$}

\author{J. H. Song, J. H. Park, K. -B. Lee, and Y. H. Jeong}

\affiliation{\it Department of Physics and electron Spin Science
Center,
 Pohang University of Science and Technology, Pohang, Kyungbuk,
790-784, S. Korea}

\begin{abstract}
Epitaxial thin films of stoichiometric ${\rm LaMnO_3}$ were grown
on ${\rm SrTiO_3(110)}$ substrates using the pulsed laser
deposition technique. From the high resolution x-ray diffraction
measurements, the lattice parameters were determined as a function
of temperature and the cooperative Jahn-Teller transition was
found to occur at $T_{JT}$=573.0 K. Also measured was resonant
x-ray scattering intensity of the orthorhombic (100) peak of ${\rm
LaMnO_3}$ near the Mn K edge from low temperatures to above
$T_{JT}$. We demonstrate that the integrated intensity of the
(100) peak is proportional to the 3/2 power of the orthorhombic
strain at all temperatures, and thus provide an experimental
evidence that the resonant scattering near the Mn K edge in ${\rm
LaMnO_3}$ is largely due to the Jahn-Teller effect.
\end{abstract}

\pacs{75.30.Vn, 78.70.Ck, 61.10.Dp, 71.90.+q}
\maketitle

Orbital degrees of freedom representing anisotropically shaped
$3d$ orbitals are of importance for understanding magnetic,
structural, and electronic properties of transition metal oxides
\cite{tokura}. For instance, superexchange interactions between
electrons occupying neighboring transition metals are very
sensitive to the orientation of occupied $3d$ orbitals
\cite{goodenough1,kanamori,khomskii}, and the magnetic properties
are strongly coupled to the orbital degrees of freedom. Also
important regarding orbital degeneracy is the well-known
Jahn-Teller (JT) distortion \cite{jahn}; moreover, these local
distortions may even behave cooperatively leading to a structural
transition. Electronic transport properties of transition metal
oxides, of course, is also strongly affected by the relative
orientation of occupied orbitals on neighboring ions; the double
exchange mechanism is a good example \cite{zener}.

Despite a recognized role that the orbital degrees of freedom play
in transition metal oxides, securing a {\it direct} experimental
probe for orbital order has long been problematic; however,
Murakami {\it et al.} proposed in a recent series of experimental
works that resonant x-ray scattering (RXS) is a direct probe of
orbital order \cite{murakami}. The RXS signal near the Mn K edge
in perovskite manganite samples was interpreted to be sensitive to
the status of $3d$ electrons and their orbital ordering. Since the
transition was supposed to be a dipolar one from $1s$ to $4p$, the
origin of the sensitivity of RXS to $3d$ orbital order became
controversial. Murakami {\it et al.} originally interpreted the
RXS data based on the localized $4p$ states and their subsequent
splitting due to the interaction with $3d$ orbitals. Ishihara {\it
et al.} then specified the interaction to be the Coulomb repulsion
between $3d$ and $4p$ states \cite{ishihara}. However, this $3d$
orbital ordering interpretation was immediately questioned by
Elfimov {\it et al.} \cite{sawatzky}; they used band calculations
to show that the JT effect (JTE) is a major player in RXS. (JTE is
used here to include both local JT distortion of ${\rm MnO_6}$
octahedra and their subsequent cooperative ordering.) Indeed it is
expected that $4p$ states form a broad band in solids, and the
$3d-4p$ direct Coulomb interactions in the band would not be as
strong as in localized states; a number of band calculations for
${\rm LaMnO_3}$ now assert that Mn $4p$ states strongly hybridize
with the neighboring O orbitals and the RXS intensity and its
polarization dependence are determined mostly by this Mn$-$O
hybridization and thus by JTE, rather than by the orbital order of
the $3d$ states \cite{sawatzky,benfatto,takahashi,benedetti}.

${\rm LaMnO_3}$, a parent material of colossal magnetoresistance
manganese oxides, has been at the center of orbital physics from
the beginning, that is, Goodenough invoked orbital order to
account for the A-type antiferromagnetic order in ${\rm LaMnO_3}$
\cite{goodenough2}. Thus ${\rm LaMnO_3}$ appears to be a good
system for the observation of orbital ordering, and Murakami {\it
et al.} measured RXS from a single crystalline ${\rm LaMnO_3}$ to
investigate orbital ordering. However, this system undergoes a
concomitant cooperative JT transition \cite{powder}, and the
characteristics of RXS can also be explained in terms of the JTE
as noted above. In the present study we attempted to address this
point experimentally by focusing specifically  on the relationship
between the lattice distortion from the JTE and the RXS intensity.
We demonstrate that the RXS signal is in proportion to the 3/2
power of the lattice orthorhombicity in ${\rm LaMnO_3}$ at all
temperatures; this result provides an evidence that the long range
order of JT distorted MnO$_6$ octahedra and the associated band
effect are indeed a dominant factor in RXS.

The samples used in the present experiment were {\it epitaxial}
thin films of ${\rm LaMnO_3}$ deposited on ${\rm SrTiO_3}$
substrates. Using a ${\rm LaMnO_3}$ bulk pellet as a target, thin
films of thickness approximately 2000 \AA\  were deposited on
${\rm SrTiO_3}(110)$ single crystal substrates by the pulsed laser
deposition method \cite{koo}. The substrates, of typical dimension
5 mm x 5 mm, were annealed for approximately 10 hours at ${\rm
1000^oC}$ in oxygen environment to obtain atomically flat
surfaces, and  then films were deposited on the substrates at
${\rm 800^oC}$ and 180 mTorr of oxygen pressure. Since as-grown
${\rm LaMnO_3}$ films may contain excessive oxygen,
post-deposition thermal treatment was done  at about ${\rm
950^oC}$ in flowing argon. The post-annealing procedure also
improved the structural quality of the films; the mosaic width
(FWHM) of the (200) reflection was about ${\rm 0.2^o}$. X-ray
scattering measurements were performed on the beamline 3C2 at the
Pohang Light Source. The incident beam is focused by a bent mirror
and is monochromatized by a Si(111) double crystal. The energy was
calibrated using a manganese foil and the resolution was about 2
eV.

At this point, remarks on the choice of the substrate and film
orientation are in order. It is known from the neutron study of
bulk samples that ${\rm LaMnO_3}$ below the structural transition
temperature (ca. 750 K) has an orthorhombic $Pbnm$ crystal
structure, and  the orthorhombic unit cell volume is $\sqrt{2}$ x
$\sqrt{2}$ x 2 times as large as that of the perovskite cubic cell
\cite{powder}. We designate these directions and lattice
parameters as $a$, $b$, and $c$, and the parameters satisfy the
relation, $a>b>c/\sqrt{2}$. (The Miller indices used for ${\rm
LaMnO_3}$ in this paper are orthorhombic ones.) Neutron powder
diffraction also indicated that the JT distortion of MnO$_6$
octahedra accompanies the structural change from a high $T$
pseudocubic phase to a low $T$ orthorhombic one, and thus the
transition is a cooperative JT type. In the orthorhombic phase,
antiferro-type $3d$ orbital ordering  and concomitant ordering of
the distorted octahedra occur in the $ab$-plane
\cite{goodenough2}. Thus in order to investigate these ordered
states in the $ab$-plane, one must probe the normally forbidden
$(h00)$ or $(0k0)$ peaks with odd $h$ and $k$. From this
consideration, it is preferred to have $a$- or $b$-direction as a
surface normal; one should deposit ${\rm LaMnO_3}$ on the (110)
surface of a cubic substrate, rather than on the usual (100)
surface. Then the surface-normal {\it cubic} [110] direction of
the substrate would corresponds to the {\it orthorhombic} [100]
direction of the film. ${\rm SrTiO_3}$ was chosen as substrate,
because its lattice constant (3.905 \AA) closely matches the
lattice spacing of ${\rm LaMnO_3}$.

Fig. 1 is the contour plot of x-ray diffraction data obtained at
6.529 keV of photon energy. It shows the reciprocal space mapping
in the vicinity of the substrate (200) reflection at two
temperatures. It is seen from Fig. 1(a) that the (220) peak of the
${\rm LaMnO_3}$ film (in orthorhombic notation) is in registry
with the cubic (200) peak of ${\rm SrTiO_3}$ at $T$ = 573.2 K. The
lattice spacings of the ${\rm LaMnO_3}$ film at this temperature
is nearly equal (3.928{\rm{\AA}}, 3.917{\rm{\AA}}, and
3.912{\rm{\AA}}), and the system is pseudocubic. At room
temperature, on the other hand, we find two diffraction peaks,
instead of one, from the film as plotted in Fig. 1(b). This
indicates that one has two domains of different orientations,
denoted as $A$ and $B$ domains (corresponding to $F_a$ and $F_b$
in the figure, respectively). Clearly this is the result of the
effort of the system which tried to reduce the strain energy as
much as possible when the system underwent the transition from a
pseudocubic to orthorhombic phase. Domain formation is often found
in a thin film which undergoes a structural change \cite{domain}.
The difference between the two domains lies in their relative
orientation, and we focus on the $A$ domain only. The lattice
constants and orientation of the $A$ domain were determined from
in-plane and out-of-plane reflections. The lattice constants in
terms of the cubic cell parameters were 3.983{\rm{\AA}},
3.912{\rm{\AA}}, and 3.866{\rm{\AA}}; the domain had its $c$-axis
(with the shortest lattice spacing) parallel to [001] in the
substrate surface, and the $a$-axis parallel to the substrate
surface normal [110]. Thus the $ab$-planes of the $A$ domain stood
perpendicular to the substrate surface.

In order to characterize the phase transition, we carried out
x-ray diffraction measurements as a function of temperature. In
Fig. 2 plotted are $a$ and $b$ lattice constants and the
orthorhombic strain $(a-b)$ against temperature, and it is seen
that the structural transition occurs at $T_{JT}$ = 573.0 K. This
transition, of course, corresponds to the cooperative JT one which
occurred at 750 K in bulk ceramic samples. This reduction in
transition temperature is not at all surprising, considering the
fact that a thin film is generally under biaxial stress due to
lattice mismatch with the substrate and/or differential thermal
expansion between the film and substrate. The fact that a single
crystal bulk sample shows the transition at 780 K \cite{murakami},
30 K higher than the ceramic case, clearly illustrates the role of
stress in determining the transition temperature. If the lattice
constants of the epitaxial film at room temperature (given above)
are compared with those of a bulk sample at the same temperature
(4.060{\rm{\AA}}, 3.912{\rm{\AA}}, and 3.834{\rm{\AA}})
\cite{elemans}, it is recognized that the orthorhombicity of the
film is much less than that of the bulk. Thus the underlying ${\rm
SrTiO_3}$ substrate has the effect on the film of suppressing the
JT transition.  The decrease of the transition temperature in thin
film case is also in accord with a recent pressure study of ${\rm
LaMnO_3}$ that the coherent JT distortion is reduced with
increasing pressure \cite{loa}.

Having characterized the cooperative JT transition, we turn to the
discussion of RXS near the Mn K edge; here we focus on the
structurally forbidden (100) peak of ${\rm LaMnO_3}$. Note that
the momentum transfer direction for this peak of the $A$ domain is
the surface normal. Fig. 3 displays the characteristics of
absorption and RXS obtained at 15.5 K from the (100) peak of the
$A$ domain. Fig. 3(a) is the plot of fluorescence data of
Mn$^{3+}$ K absorption as a function of incident photon energy.
The shape of the curve including the splitting into two peaks
matches the $4p$ total density of states from the band
calculation, indicating the dipolar nature of the process
($1s\rightarrow 4p$) \cite{sawatzky}. The (100) peak exhibits a
resonance-type enhancement in the integrated intensity as a
function of energy and reaches a maximum at 6.555 keV as
illustrated in Fig. 3(b). Since the incident synchrotron beam is
$\sigma$-polarized (polarization perpendicular to the scattering
plane) and RXS occurs via the $\sigma\rightarrow\pi$ channel, the
RXS intensity should exhibit an oscillation in intensity as the
$ab$-plane is rotated around the momentum transfer direction,
i.e., the surface normal. Fig. 3(c) shows the results of the
azimuthal scan measured at 6.555 keV of energy, revealing a
sinusoidal variation with a two-fold symmetry. The integrated
intensity plotted in the figure was normalized with the (200)
fundamental charge reflection in order to correct for small
variations due to the film shape. The azimuthal angle $\psi=0^o$
was defined to be when the $ab$-plane is perpendicular to the
scattering plane.

The features displayed in Fig. 3, resonance enhancement of the
(100) peak and the two-fold symmetric nature of the integrated
intensity as a function of azimuthal angle, follow from the fact
that RXS is a second order process driven by the
$\vec{A}\cdot\vec{p}\ $  term in the Hamiltonian; however, the
same result can be derived either from the antiferro-type $3d$
orbital ordering or from the coherent ordering of the JT distorted
${\rm MnO_6}$ octahedra in the $ab$-plane. The essential
ingredient in the theory is a breaking of the equivalence of two
Mn sites (and the associated non-spherical $4p$ states) in the
orthorhombic $ab$-plane, and this can be done by either of the two
ways. Since these two types of order coexist in ${\rm LaMnO_3}$,
it appears to be difficult to distinguish them as a origin of RXS.
On the other hand, if one remembers that coherent ordering of
local JT distortions brings about the structural transition, while
orbital ordering itself seems to occur even without lattice
distortion \cite{endoh}, then one is led to a conclusion: if the
JTE is indeed the cause of RXS in ${\rm LaMnO_3}$, there {\em
must} exist a certain relationship between the intensity of RXS
and the degree of lattice distortion which would be proportional
to the degree of local JT distortion. Thus, we focused our
attention on revealing this relationship from the experimental
data; the results are illustrated in Fig. 4.

Fig. 4(a) is the plot of the integrated intensity of the (100)
peak as a function of temperature. The integrated intensity was
measured at 6.555 keV of photon energy and the azimuthal angle was
set at $\psi=0^o$.  The intensity disappears exactly at the same
temperature $T_{JT}$ where the orthorhombic strain does, as
temperature is increased. Fig. 4(b) is the simultaneous plot of
the integrated intensity and the normalized orthorhombic strain $e
\equiv (a-b)/(a+b)$ as a function of temperature. Note that the
vertical axis for the normalized  strain is not $e$, but
$e^{3/2}$. The fact that these two data sets can be superposed at
all temperatures means that the RXS intensity is proportional to
$e^{3/2}$; the inset, the plot of $e^{3/2}$ against the intensity,
further highlights the proportional relationship. This simple
relationship we exposed here, particularly the nontrivial exponent
3/2, must be able to distinguish the two proposed origins of RXS.
In fact, the ab initio band calculation on ${\rm LaMnO_3}$ by
Benedetti {\it et al.} \cite{benedetti}, which identified the JTE
as the origin of RXS near the Mn K edge, predicted the
proportional relationship between the integrated instensity and
$e^{1.5}$. The present work then should be regarded as a definite
experimental evidence that the resonant scattering near the Mn K
edge in ${\rm LaMnO_3}$ is due to the Jahn-Teller effect, rather
than $3d$ orbital ordering.

One final remark on the relationship between the lattice
orthorhombicity and the local JT distortion should be made,
because one of these two aspects of ${\rm LaMnO_3}$ does not
automatically suggest the other.  The orthorhombic symmetry may be
induced from cubic perovskites just by a tilting of regular, not
distorted, oxygen octahedra, e.g., ${\rm CaTiO_3}$ \cite{cto}. In
fact, even the high $T$ phase of ${\rm LaMnO_3}$, where oxygen
octahedra are regular, belongs to the $Pbnm$ class, not a cubic
one \cite{powder}. These systems, however, are in sharp contrast
to ${\rm LaMnO_3}$ below $T_{JT}$, that is, their three lattice
parameters are very close in size to each other despite
orthorhombicity, and they may be called {\em pseudocubic}. The
significant lattice orthorhombicity in low $T$ ${\rm LaMnO_3}$ is
the direct result of the local JT distortion: the long and short
Mn-O bonds alternate in the $ab$-plane and contribute to $a$ and
$b$ lattice parameters in different proportions due to the tilting
of octahedra \cite{argyriou}. It is also noted that both the
difference $(a-b)$ of the lattice parameters and the difference in
Mn-O bond lengths disappear simultaneously and continuously as $T$
is increased toward $T_{JT}$. Evidently the lattice
orthorhombicity and the degree of local JT distortion are
proportional to each other.

In conclusion, we have demonstrated that the RXS intensity  is
proportional to the 3/2 power of the orthorhombic strain, and
thereby showed that RXS near the Mn K edge is due to the JTE,
rather than $3d$ orbital ordering.

We acknowledge the financial support from KOSEF via eSSC, the
special fund of POSTECH, and the BK21. Experiments at the PLS were
supported by MOST and POSCO. We thank Y.J. Park for technical
assistance.

\newpage

{\bf\centering Figure Captions}\\

\noindent FIG. 1: The contour plot of x-ray diffraction data
around the cubic (200) reflection of ${\rm SrTiO_3}(110)$.
$q_{\bot}$ and $q_{\|}$ are parallel to the [110] and
[1$\bar{1}$0] directions, respectively. (a) At 573.2 K, a single
peak in registry with the substrate (200) peak is found from the
film. Inset illustrates the momentum transfer of the substrate
(200) peak. $F$ and $S$ denote the peaks from the film and
substrate, respectively. (b) At room temperature, the two peaks
(denoted as $F_a$ and $F_b$) from two domains with different
orientation are seen. Inset shows
diffraction data along the surface-normal direction.\\

\noindent FIG. 2: The lattice constants $a$ and $b$  and the
orthorhombic strain $(a-b)$ as a function of temperature. The
cooperative Jahn-Teller transition occurs at $T_{JT}$ = 573.0 K.
This transition temperature is somewhat low compared to 750 K of a
bulk sample. This is due to the fact that the film is under
biaxial
stress. The lines are guide to the eye.\\

\noindent FIG. 3: X-ray absorption and resonant scattering near
the Mn K edge. The data were obtained from the structurally
forbidden (100) peak of the $A$ domain at 15.5 K. (a) The
fluorescence data is plotted as a function of photon energy. The
shape of the curve matches well the $4p$ total density of states
from the band calculation, indicating the $1s\rightarrow 4p$
process. (b) The integrated intensity for the (100) peak as a
function of energy. The intensity exhibits a resonance-type
enhancement around a maximum at 6.555 keV. (c) The integrated
intensity of the (100) peak measured as a function of azimuthal
angle is shown. The solid
line represents a fit of the data to $\cos^2 \psi$.\\

\noindent FIG. 4: (a) Temperature dependence of the integrated
intensity for the (100) peak. The intensity disappears at $T_{JT}$
where the orthorhombic strain does, as temperature is increased.
(b) The integrated intensity and the 3/2 power of the normalized
orthorhombic strain $e = (a-b)/(a+b)$ are plotted as a function of
temperature simultaneously. The complete coincidence of these two
data sets at all temperatures is further highlighted in the inset
by plotting $e^{3/2}$ vs. intensity.

\newpage

\bfig \includegraphics[width=16cm]{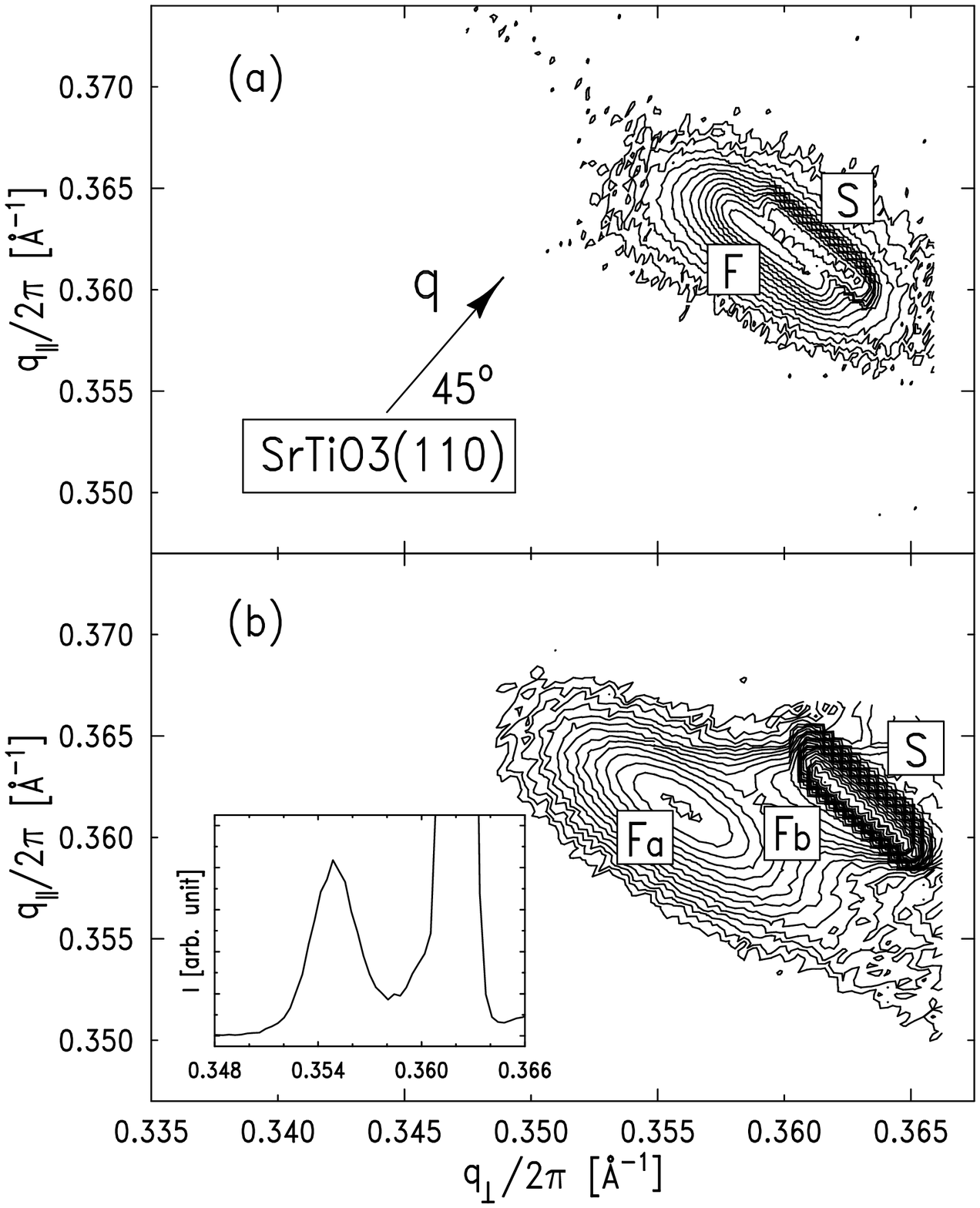} \caption{Song et al.}
\efig

\bfig \includegraphics[width=16cm]{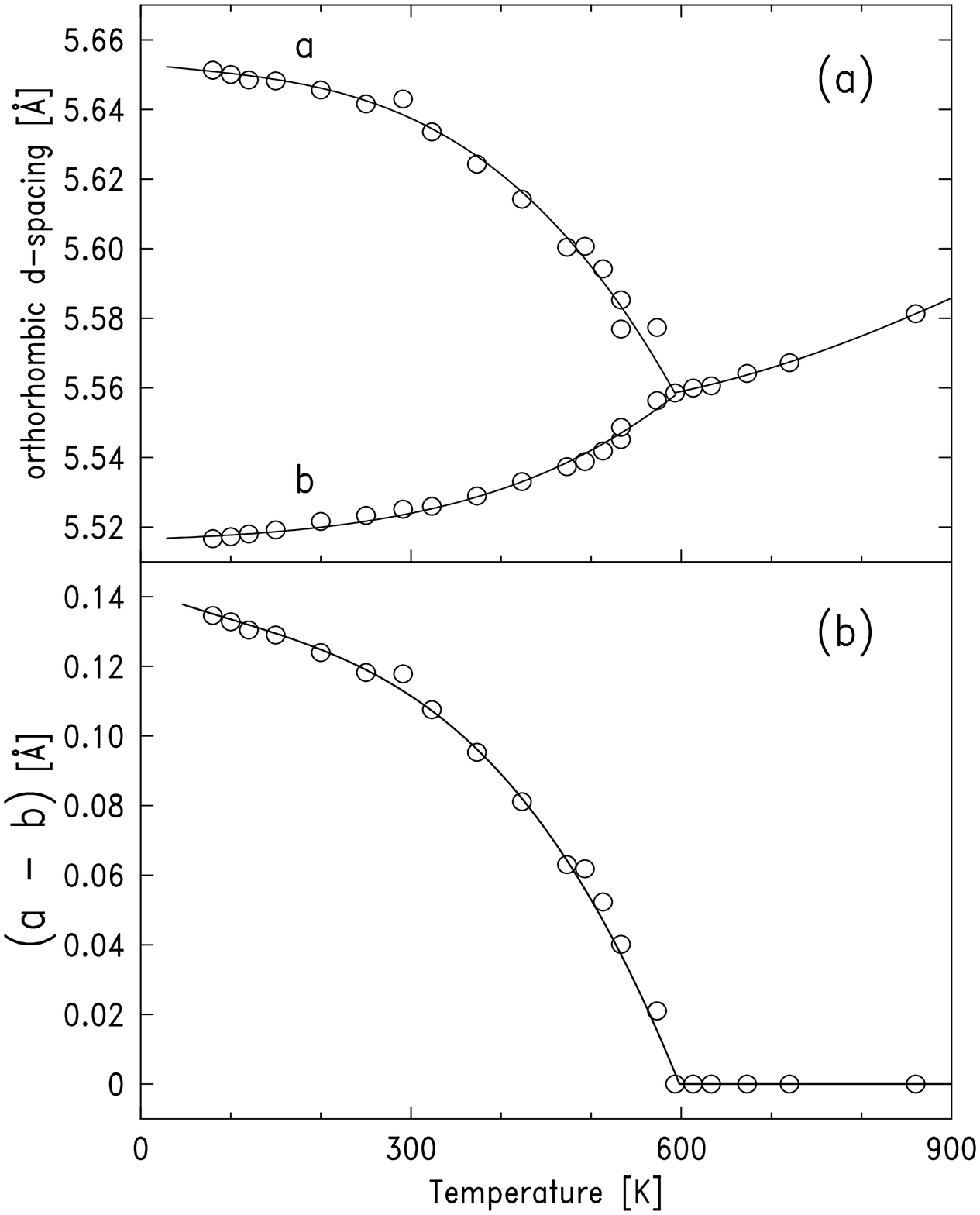} \caption{Song et al.
} \efig

\bfig \includegraphics[width=16cm]{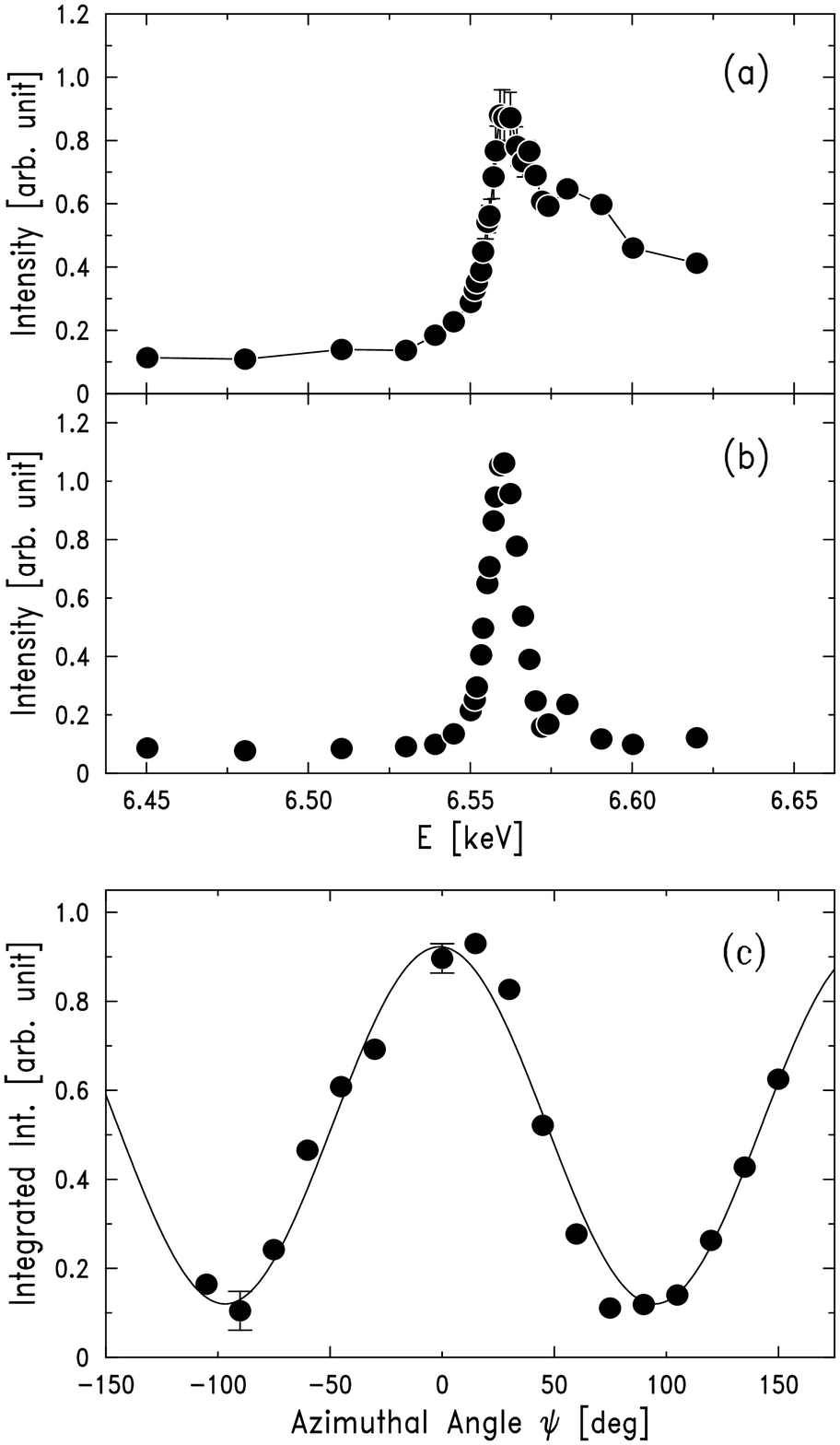} \caption{Song et al.
}\efig

\bfig \includegraphics[width=16cm]{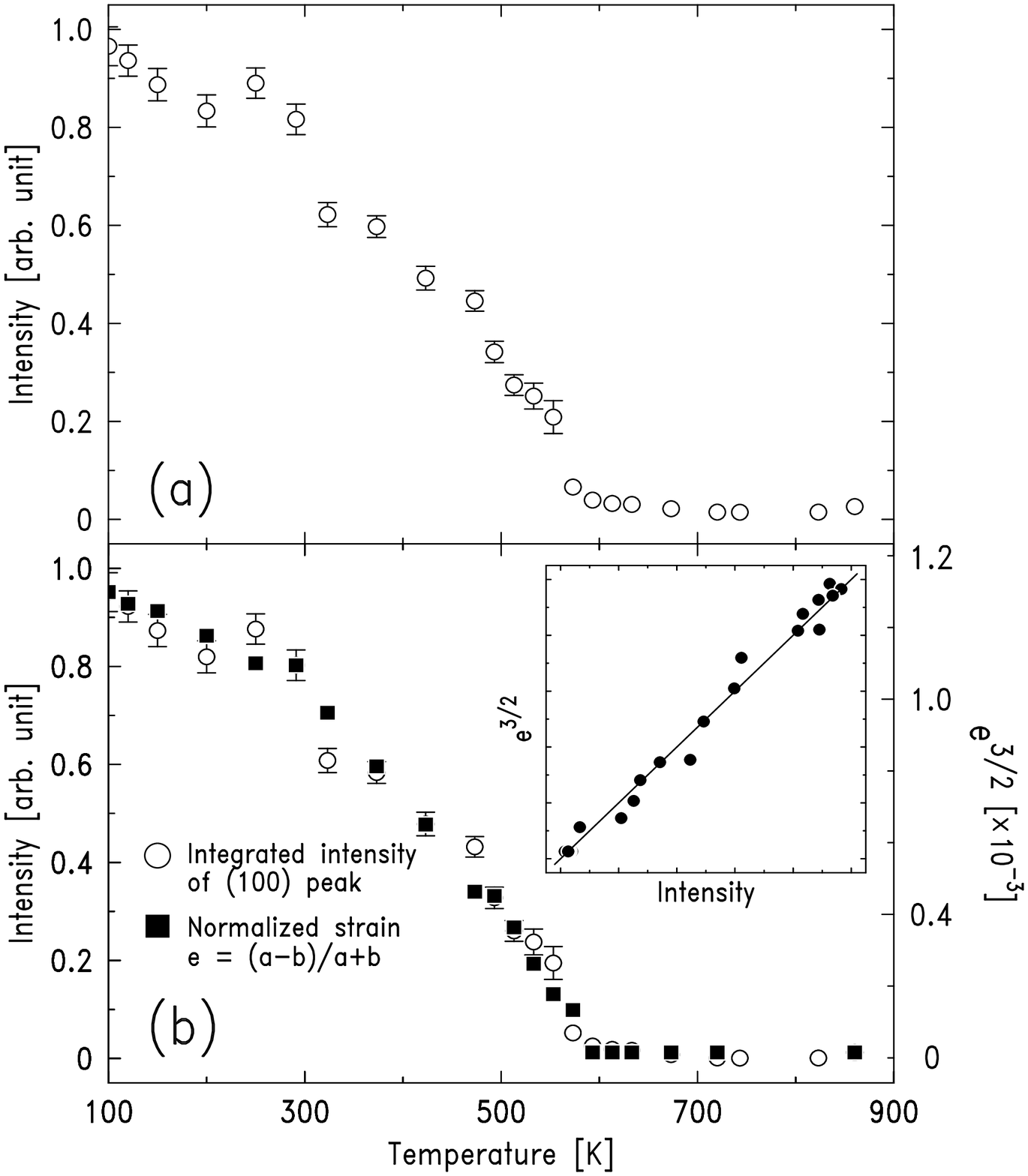} \caption{Song et al.
}\efig

\end{document}